\documentclass[%
twocolumn,
reprint,
superscriptaddress,
 amsmath,amssymb,
 aip,
 jcp,
]{revtex4-1}
\reversemarginpar

\newcommand{\coo}{CO$_2$}
\newcommand{\etabs}{\eta_{\rm b}/\eta_{\rm s}}
\newcommand{\be}{\begin{equation}}
\newcommand{\ee}{\end{equation}}
\newcommand{\benn}{\begin{displaymath}}
\newcommand{\eenn}{\end{displaymath}}
\newcommand{\sfrac}[2]{\mbox{$\frac{#1}{#2}$}}

\usepackage{graphicx,tabularx}
\usepackage{xspace}
\usepackage{dcolumn}
\usepackage{bm}
\usepackage[version=3]{mhchem}
\usepackage{verbatim}
\usepackage{hyperref}
\usepackage{cleveref}
\usepackage{amsmath}
\usepackage{amssymb}
\usepackage{multirow}
\usepackage{natbib}

\begin{document}

\title{Bulk viscosity of CO$_2$ from Rayleigh-Brillouin light
scattering spectroscopy at 532 nm}

\author{Yuanqing Wang}
\author{Wim Ubachs}
\email{w.m.g.ubachs@vu.nl}
\affiliation{Department of Physics and Astronomy, LaserLaB, Vrije Universiteit, De Boelelaan 1081, 1081 HV Amsterdam, The Netherlands}

\author{Willem van de Water}
\affiliation{Laboratory for Aero and Hydrodynamics, Faculty of
Mechanical, Maritime and Materials Engineering, Delft University of
Technology, Leeghwaterstraat 29, 2628CB Delft, The Netherlands}

%
%
%
\date{\today}

\begin{abstract}
%

Rayleigh-Brillouin scattering spectra of \coo\ were measured at
pressures ranging from 0.5 to 4~bar, and temperatures from 257 to
355~K using green laser light (wavelength 532~nm, scattering angle of 55.7$^\circ$).  These spectra
were compared to two lineshape models, which take the bulk viscosity
as a parameter.  One model applies to the kinetic regime, i.e. low
pressures, while the second model uses the continuum,
hydrodynamic approach and takes the rotational relaxation time as a
parameter, which translates into the bulk viscosity. We do
not find a significant dependence of the bulk viscosity with pressure
or temperature.  At pressures where both models apply we find a
consistent value of the ratio of bulk viscosity over shear viscosity
$\etabs = 0.41 \pm 0.10$.  This value is four orders of magnitude
smaller than the common value that is based on the damping of
ultrasound, and signifies that in light scattering only relaxation of
rotational modes matters, while vibrational modes remain 'frozen'.

\end{abstract}

\maketitle

\section{Introduction}
\label{SecCO2GreenIntroduction}

The light scattering properties of carbon dioxide remain of interest,
both from a fundamental perspective studying the relaxation in
molecular gases and for determining their thermodynamic properties,
as well as from an applied perspective. Details of Rayleigh-Brillouin
(RB) phenomena, scattering spectral profiles of CO$_2$ gas at
differing pressures and temperatures \cite{Lao1976a, Pan2005,
Gu2014a} as well as its cross section \cite{Sneep2005}, are of
relevance for current and future remote sensing exploration of the
planetary atmospheres where CO$_2$ is the main constituent, either
under high-pressure and high-temperature conditions as on Venus
\cite{Rasool1970} or under low pressure and low-temperature
conditions as on Mars \cite{Aharonson2004}. The fact that carbon
dioxide is the prime greenhouse gas has spurred large-scale activity
in the transformation of this gaseous species
\cite{Sakakura2007,Rooij2018}, through catalytic
hydrogenation~\cite{Wang2011}, electrochemical conversion into
renewable energy \cite{Kauffman2015}, as well as in plasma-driven
dissociation for the synthesis of fuel from
CO$_2$~\cite{Kondratenko2013,Bongers2017}. Apart from issues of
capture, fixation and transformation of Teratons of carbon dioxide,
its storage and transport, either in the liquid or gas phase, forms
an important challenge \cite{Mikkelsen2010,Boot-Handford2014}. For
these purposes study of the transport coefficients of CO$_2$ gas,
such as thermal conductivity \cite{Uribe1990}, heat capacity, and
shear viscosity \cite{Trengove1987,Bock2002}, is of practical
importance.
Light scattering is an elegant way to determine the
thermodynamic properties of a gas, because the Rayleigh scattering
phenomenon resulting in the elastic peak is connected to entropy
fluctuations \cite{Strutt1899}, while the Brillouin-side peaks are
associated with density fluctuations or sound
\cite{Brillouin1922,Mandelstam1926}. The macroscopic gas transport
coefficients govern the scattering spectral profiles and can in turn
be deduced from measurement of such profiles \cite{Hirschfelder1948}.
This holds for both spontaneous RB-scattering
\cite{Letamendia1982,Marques1993,Vieitez2010} as well as for coherent
RB-scattering \cite{Pan2004,Meijer2010}.

The bulk viscosity $\eta_{\rm b}$ \cite{Cramer2012,Jaeger2018}, is the most elusive
transport coefficient.  It is associated with the relaxation of
internal degrees of freedom of the molecule, i.e. rotations and
vibrations. The bulk viscosity is commonly measured from the damping
of ultrasound at frequencies in the MHz domain \cite{Herzfeld1959}. It can also be retrieved from the light scattering spectrum of molecular gases. This
was demonstrated by \citet{Pan2005} and recently for N$_2$, O$_2$ and
air by \citet{Gu2014b}, and for N$_2$O gas by \citet{Wang2018}.
In light scattering, the frequencies $f_{\rm s}$ involved are those of sound
with wavelengths comparable to that of light, three orders of
magnitude larger than the frequencies used to measure $\eta_{\rm b}$
from ultrasound experiments. The bulk viscosity of \coo\ from light
scattering is found to be four orders of magnitude smaller than that from ultrasound experiments\citep{Lao1976a,Pan2005,Gu2014a}. A simple explanation is that at high (hypersound)
frequencies the relaxation of vibrational modes of the \coo\ molecule
no longer plays a role, i.e. the vibrational energy stays frozen in.

The bulk viscosity can be expressed in terms of relaxation times of
intra-molecular degrees of freedom,
\be
   \eta_{\rm b} = p \frac{2}{\left(3 + \sum_i N_i\right)^2}
   \sum_i N_i \: \tau_i
\label{Eq:ModifiedEuckenSimple}
\ee
where $p$ is the pressure, $N_i$ is the number of degrees of freedom of mode $i$
(rotations, vibrations), and $\tau_i$ is the relaxation time
($\tau_{\rm rot}, \: \tau_{\rm vib}$) \cite{Chapman1970}.  A
frequency-dependent version of this formula, depending on the
product $f_{\rm s}\tau_i$ was given by \citet{Meijer2010}.  Carbon dioxide
is a linear molecule with 2 rotational degrees of freedom.  In case
of frozen vibrations ($f_{\rm s} \tau_{\rm vib} \gg 1$), Eq.\ (\ref{Eq:ModifiedEuckenSimple})
reduces to
\be
   \eta_{\rm b} = \frac{4}{25} \: p \: \tau_{\rm rot}
\label{Eq:ModifiedEuckenSimpleRelax}
\ee
The vibrational relaxation time strongly decreases with temperature,
and so does the bulk viscosity. In a simple model Landau and Teller proposed an exponential dependence of the chance of relaxation on the ratio of the collision interaction time and the vibration period
\cite{Herzfeld1959}. This leads to a scaling prediction for the
temperature dependence of the bulk viscosity which agrees with
experiment \cite{Cramer2012}. On the other hand, a classical
analysis of collisions of rigid rotators by \citet{Parker1959}
results in a scaling expression for $\tau_{\rm rot}$ which predicts an
{\em increase} of $\tau_{\rm rot}$ with increasing temperature.

That the bulk viscosity changes with temperature motivated the present study in which laser-based light
scattering measurements in CO$_2$ gas are carried out in a pressure
regime of 0.5 -- 4 bar and in a temperature regime of 257 -- 355 K.
Accurate, highly spectrally resolved and high signal-to-noise
scattering line profiles are measured at a scattering angle of 55.7
degrees and a scattering wavelength of $\lambda_i = 532$ nm, a
wavelength commonly used in Lidar applications. Under these
conditions of a longer wavelength and a smaller scattering angle than
in a previous study \cite{Gu2014a}, the Brillouin-side peaks become
more pronounced in the scattering spectrum.

Experimental data are analyzed in the context of two models for
the spectral lineshape. One model applies at low pressures, the
kinetic regime where the mean free path between collisions is
comparable to the wavelength of light \citep{Boley1972,Tenti1974},
while the other one is valid in the hydrodynamic regime
\citep{Hammond1976}. Values of the bulk viscosity are determined in a
least squares method by comparing model spectra to measured ones.

The remainder of the article consists of an experimental section, a
section discussing the bulk viscosity and model descriptions for
RB-scattering, a presentation of results in the context of
applicable models to describe the scattering spectrum, followed by a
conclusion.

\section{Experiment}
\label{SecCO2GreenExperiment}

\begin{figure*}[ht]
\centering
\includegraphics[scale=0.55]{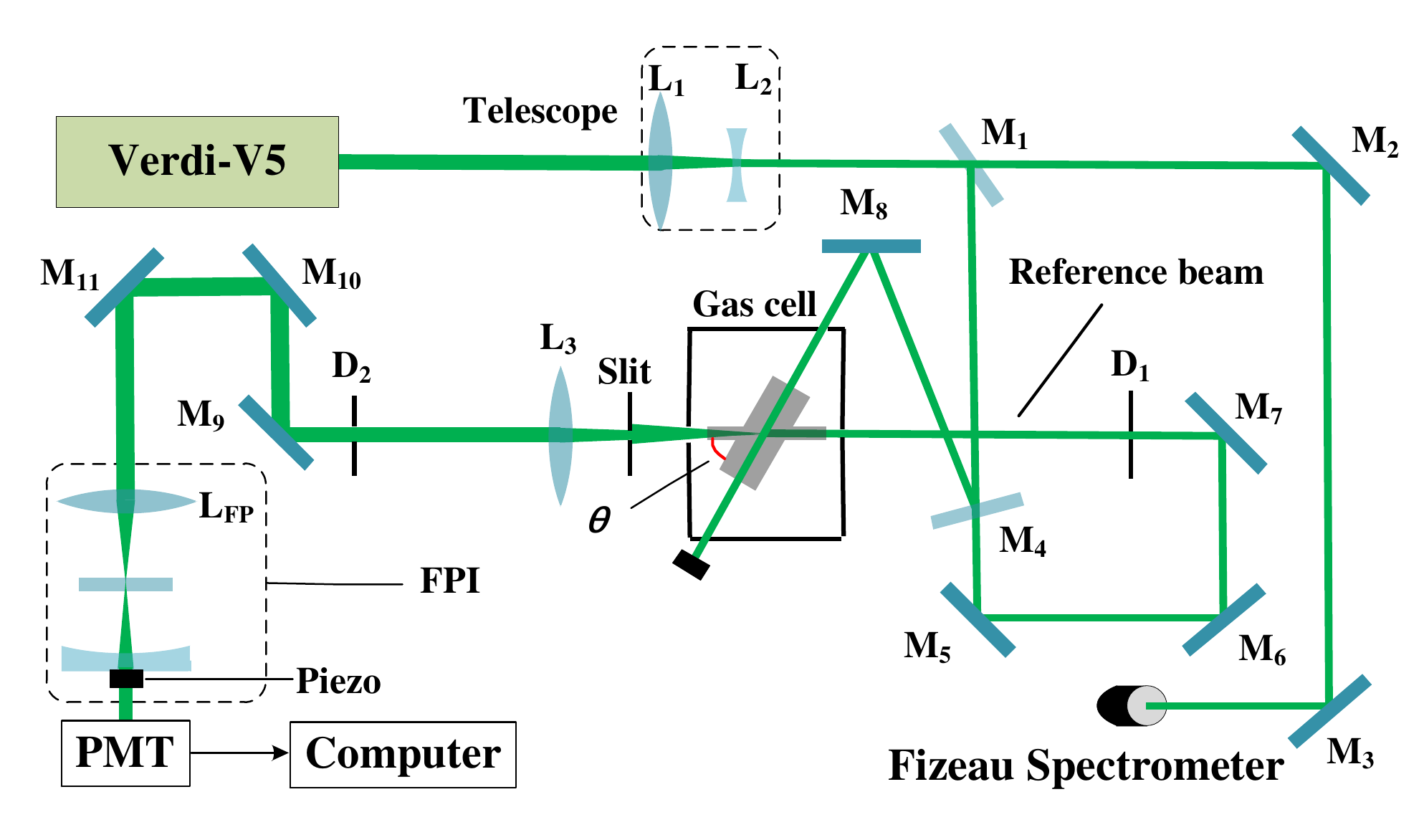}
\caption{Schematic of the experimental apparatus. A Verdi-V5 laser provides
continuous wave light at 532.22 nm at a power of 5 Watt and bandwidth
less than 5 MHz. The laser light is split into two beams: The pump
beam crosses the RB-scattering gas cell producing scattered light
that is captured under an angle $\theta =(55.7 \pm 0.3)^{\circ}$. The
small fraction reference beam transmitted through M$_1$ is used to align
the beam path after the gas cell towards the detector. The scattered
light is analyzed in a Fabry-Perot Interferometer(FPI), with  free
spectral range of 2.9964 GHz and an instrument linewidth of ($58 \pm 3$) MHz, and is
collected on a photo-multiplier tube (PMT).  Mirrors, lenses and
diaphragm pinholes are indicated as M$_i$, L$_i$ and D$_i$. A slit of
500 $\mu$m is inserted to limit the opening angle for collected
scattering light, therewith optimizing the resolution.
}
\label{Fig:SetupGreen}
\end{figure*}

The experimental apparatus for measuring the RB-scattering spectral profiles of CO$_2$ is displayed in
Fig.\ ~\ref{Fig:SetupGreen}. The laser source provides continuous wave
radiation at $\lambda_i = 532.22$ nm at a bandwidth of less than 5
MHz. RB-scattered light is produced from the laser
beam of 5 Watt intensity traversing a gas cell equipped with a gas
inlet valve and a pressure sensor. Brewster-angled windows are
mounted at entrance and exit ports, and black paint covers the inside
walls, to reduce unwanted scatter contributions. A temperature
control system equipped with Peltier elements is employed for
heating, cooling and keeping the cell at a constant temperature with
uncertainty less than $0.1 \: ^\circ{\rm C}$. RB-scattered light is
captured at a scattering angle $\theta = (55.7 \pm 0.3)^\circ$
defined by the setting of a slit in the scatter beam path (see
Fig.\ \ref{Fig:SetupGreen}), that also limits the opening angle for
collecting RB-scattered light to less than 0.5$^\circ$. The exact
scatter angle and uncertainty are determined with a rotatable stage
operated as a goniometer. A reference beam as depicted in
Fig.\ \ref{Fig:SetupGreen} is used for aligning the collection and
detection system.  The scattering angle determines the scattered
light wave vector $k_{\rm sc}$,
\benn
   k_{\rm sc} = \frac{2\pi}{\lambda_i}\: 2 \: n \: \sin(\theta/2)
\eenn
with $n$ the index of refraction.

The scattered light propagates through a bandpass filter (Materion, T
$>$ 90$\%$ at $\lambda_i$ = 532 nm, bandwidth $\triangle\lambda$ = 2.0
nm) onto a Fabry-Perot interferometer (FPI) via an optical projection
system consisting of a number of lenses and pinholes to reduce stray
light and contributions from Raman scattering. Finally the scattered
photons are detected on a photomultiplier tube (PMT), processed and
stored in a data acquisition system.
The FPI is half-confocal, the curved mirror having a radius of
curvature of $r = -12.5$ mm. Mirror reflectivities are 99\%. The FPI
has an effective free spectral range (FSR) of $2.9964$ GHz, which is
determined through frequency-scanning a laser (a narrowband tunable
cw-ring dye laser) over more than 1000 modes of the FPI while
measuring the laser wavelength by a wavelength meter (Toptica
HighFinesse WSU-30), hence yielding an uncertainty in the FSR below 1
MHz. The instrument width, yielding a value of $\sigma_{\nu_{\rm
instr}}$ = 58.0 $\pm$ 3.0 MHz (FWHM), is determined by using the
reference beam while scanning the piezo-actuated FPI, following
methods discussed by \citet{Gu2012rsi}. It includes the bandwidth of
the incident laser. The instrument function is verified to exhibit
the functional form of an Airy function, which may be well
approximated by a Lorentzian function during data analysis.

\begin{figure*}[t]
\centering
\includegraphics[scale=0.4]{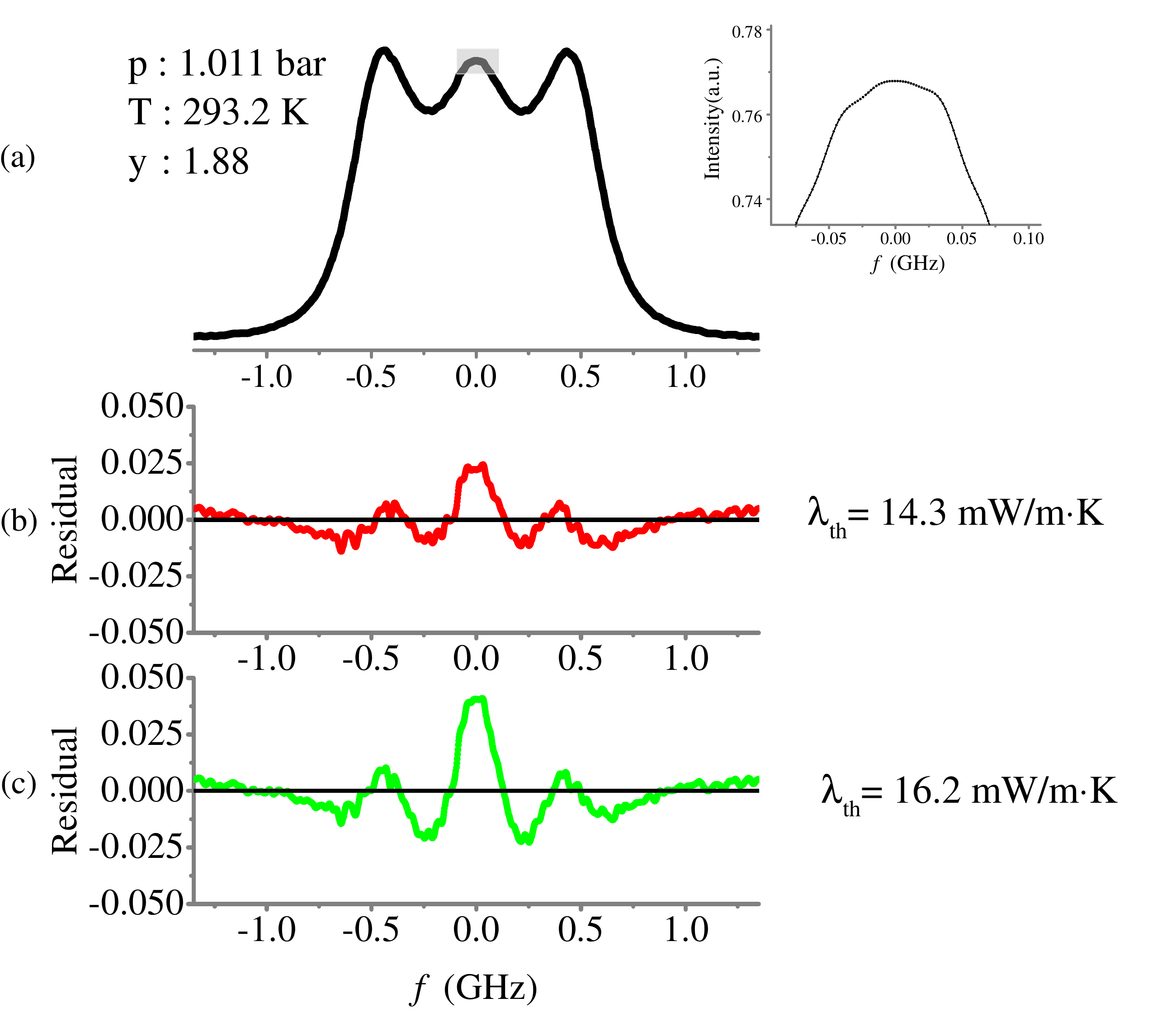}
\caption{(a) Experimental data for the RB-light scattering spectrum of
   CO$_2$ measured at $\lambda_i=532$ nm, $\theta=55.7^{\circ}$
   and ($p$,$T$) conditions as indicated, corresponding to a
   uniformity parameter $y = 1.88$; the spectrum on the right shows
   en enlargement of the central part indicated in grey; (b)
   Residuals from comparison with Tenti-S6 model by using thermal
   conductivity $\lambda_{\rm th}$ from Eq.\ (\ref{Eq:ModifiedEucken}) and a
   fitted value for $\eta_{\rm b}$; (c) Residuals with $\lambda_{\rm th}$ from
   \citet{Uribe1990}. }
\label{Fig:CO2GreenCompareDifferentThermal}
\end{figure*}

RB-scattering spectral profiles were recorded by piezo-scanning the
FPI at integration times of 1 s for each step, usually over 18 MHz.
Typical detection rates were $\sim 2000$ count/s for conditions of 1
bar pressure. A full spectrum covering lots of consecutive RB-peaks and
10,000 data points was obtained in about 3 h. The piezo-voltage scans
were linearized and converted to frequency scale by fitting the
RB-peak separations to the calibrated FSR-value. The linearization
procedure also corrects for frequency drifts of the laser, which were
measured to amount to $10 - 100$ MHz/hour, depending on temperature
drifts in the laboratory. Finally, a collocated spectrum was obtained
by cutting and adding all individual recordings over $\sim 60$
RB-peaks~\cite{Gu2012rsi}.  In a final step the RB scattering
profiles were averaged to improve the signal to noise ratio. This
procedure yields a noise level of $\sim 0.4$\% (with respect to peak
height) for the 1 bar pressure case. A single typical light
scattering spectrum recorded at 1 bar and room temperature, measured
in a typical recording time of $\sim 3$ h, is displayed in
Fig.\ \ref{Fig:CO2GreenCompareDifferentThermal}. This figure and its
inset demonstrate the signal-to-noise ratio attainable in the present
setup.

%

\section{RB-scattering and line shape models}
\label{models}

In light scattering the key quantity is the uniformity parameter $y$,
which -up to a constant- is defined as the ratio of the scattering
wavelength over the mean free path  between collisions, which can be shown to equal
\benn
   y =  \frac{p}{k_{\rm sc} \: v_0 \: \eta_{\rm s}}
\eenn
with thermal velocity $v_0 = (2 k_{\rm B} T / m)^{1/2}$, where
$k_{\rm B}$ is the Boltzmann constant, $m$ the molecular mass and $\eta_{\rm s}$ the shear viscosity.

Values $y = {\cal O}(1)$ pertain to the kinetic regime, and
models must be based on the Boltzmann equation.  There, spectra do
not deviate strongly from the Rayleigh (Maxwellian) line shape. At
larger values of $y$ many mean free paths fit in a wavelength, and a
hydrodynamic, continuum, approach applies.  The Brillouin-side
features become more and more prominent with increasing $y$, and
occur at frequency shifts $f_{\rm s} = \pm \upsilon_{\rm
s}k_{\rm{sc}}/2\pi$, with $\upsilon_{\rm s}$ the speed of sound. Our
data are in the interval $y = [0.7 - 9]$, and, therefore range from
the kinetic into the hydrodynamic regime.

\subsection{The Tenti model}
The Tenti model, originally developed for analyzing RB-scattering in
molecular hydrogen and diatomic molecules~\cite{Boley1972,Tenti1974},
is a widely used model for light scattering spectra in the kinetic
regime. It uses the Wang Chang--Uhlenbeck eigentheory which takes
known values of the macroscopic transport coefficients as input
\cite{Wang1951}.  This input consist of values for the shear
viscosity $\eta_{\rm s}$, thermal conductivity $\lambda_{\rm th}$, the molar
heat capacity $C_{\rm int}$ of internal modes of motion (rotations,
vibrations), and the bulk viscosity $\eta_{\rm b}$.
The model agrees well with experiment
\cite{Pan2004,Vieitez2010,Gu2014a}, where it was established that the
six-mode version of the Tenti model (hereafter called the Tenti-S6 model)  yields a better agreement with
experiment than the seven-mode variant.

We use the values for a temperature dependent shear viscosity $\eta_{\rm s}(T)$
for CO$_2$ from \citet{Bousheri1987}. Of the transport coefficients needed in the model, $\lambda_{\rm th}$
and $C_{\rm int}$ and $\eta_{\rm b}$ depend on the participation of
intra-molecular modes of motion.  We assume that at the frequencies
associated with light scattering, only rotations participate in the
exchance of internal and kinetic energy, so that $C_{\rm int}$, for
the linear \coo\ molecule becomes $C_{\rm int} = 2/2 \: R$ with $R$ the
gas constant.

For the thermal conductivity $\lambda_{\rm th}$, \citet{Uribe1990} list temperature-dependent values. As in the present study only the thermal conductivity associated with rotational relaxation is considered, these values are not straightforwardly applicable to RB-scattering data. We need a value that reflects rotational internal energy only. For polyatomic gases, a high-frequency value for $\lambda_{\rm th}$ was estimated
from the Eucken relation, which expresses $\lambda_{\rm th}$ as a function of the
shear viscosity $\eta_{\rm s}$, the diffusivity $D$ and the heat capacity $C_{\rm
int}$ of internal motion:
\be
   \lambda_{\rm th} = \sfrac{5}{2} \eta_{\rm s} \: C_{\rm t} +
   \rho \: D \: C_{\rm int}
\label{Eq:ModifiedEucken}
\ee
with $C_{\rm t} = 3 / 2 \: R$, the heat capacity of kinetic
motion, $C_{\rm int} = 2 / 2 \: R$, and the temperature-dependent
diffusivity $D$ taken from \citet{Bousheri1987}.
At temperature $T = 296.55 \: {\rm K}$, the low-frequency value is
$\lambda_{\rm th} = 1.651 \times 10^{-2}$ W/m$\cdot$K (Ref.\ \onlinecite{Uribe1990}), whereas Eq.\ (\ref{Eq:ModifiedEucken}) predicts a high-frequency value $\lambda_{\rm th} = 1.452 \times 10^{-2}$  W/m$\cdot$K.  As will be
demonstrated below, the smaller value indeed produces a better fit
of the kinetic model.
The bulk viscosity $\eta_{\rm b}$, our prime quantity of interest, is
determined in a least squares procedure.

\subsection{The Hammond--Wiggins model}
At the other end of the uniformity scale we seek confrontation with a hydrodynamic, continuum, model by \citet{Hammond1976}. Unlike the Tenti model, which is built on (tensorial) eigenvectors of the linearized collision operator, the hydrodynamic model is built on
(tensorial) moments of the space-time distribution function, i.e.
hydrodynamic quantities.
The hydrodynamic model takes the shear viscosity $\eta_{\rm s}$, the
diffusivity $D$ and the heat capacity of internal motion $C_{\rm
int}$ as parameters.  The rotational relaxation time
$\tau_{\rm rot}$ is determined in a least squares procedure, from
which the bulk viscosity $\eta_{\rm b}$ is computed using Eq.\
(\ref{Eq:ModifiedEuckenSimpleRelax}). Allowance for rotational relaxation only is done
through the choice for $C_{\rm int} = 2/2 \: R$.

The evaluation of both kinetic and hydrodynamic models can be done extremely quickly. Where their range of validity overlaps, the derived values of the bulk viscosity should agree.
%
%

\section{Results and Discussion}
\label{SecCO2GreenResults}

In this section we will first present experimental data on light
scattering in CO$_2$, followed by an analysis in terms of two
complementary spectral line models. We finally summarize the results
of the temperature-dependent bulk viscosity.
\begin{table*}[t]
{\caption{
\label{TableCO2GreenConditionsandResults}
Data sets for RB-scattering measurements in CO$_2$ gas recorded
under conditions as indicated. The uniformity parameter is $y$. For
values of the temperature-dependent $\eta_{\rm s}$ and $\lambda_{\rm
th}$ we use \citet{Bousheri1987} and Eq.\ (\ref{Eq:ModifiedEucken}), respectively.  The bulk viscosity and the ratios
$\eta_{\rm b}/\eta_{\rm s}$ are derived in a fit to the experimental
data.  The bulk viscosity $\eta^{T}_{\rm b}$ is based on the
Tenti-S6 model, while $\eta^{H}_{\rm b}$ is based on the
Hammond--Wiggins hydrodynamic model.  The parameter $C_{\rm int} = 2/2 \:
R$, for all cases is the heat capacity of rotational motion.
}}
\begin{center}
\begin{tabular}{c c c c c c c c c c}
\hline
Data set & $p$  & $T$  &$\eta_{\rm s}$($\times 10^{-5}$) &$\lambda_{\rm th}$($\times 10^{-3}$) & $\eta^{T}_{\rm b}$($\times 10^{-5}$)& $\eta_{\rm b}^{T}/\eta_{\rm s}$ &$\eta^{H}_{\rm b}$($\times 10^{-5}$) &$\eta_{\rm b}^{H}/\eta_{\rm s}$&$y$ \\
  Unit   &   bar   &   K   & Pa$\cdot$s &  W/m$\cdot$K   & Pa$\cdot$s&&Pa$\cdot$s&\\
 \hline
 \hline
  \multirow{6}*{~0.5 bar}
 &0.500&273.2&1.37&13.4&0.64 &0.47&&&1.03\\
 &0.500&293.2&1.47&14.3&0.72&0.49&&&0.93\\
 &0.505&313.2&1.56&15.3&0.91 &0.58&&&0.85\\
 &0.508&333.2&1.66&16.2&1.65 &0.99&&&0.78\\
 &0.503&353.2&1.75&17.1&1.79 &1.02&&&0.71\\
  \hline
 \multirow{7}*{~1 bar}
 &1.033&258.1&1.30&12.7&0.50 &0.38&0.30&0.23&2.32\\
 &1.038&274.3&1.38&13.4& 0.46 &0.33&0.24&0.18&2.13\\
 &1.011&293.2&1.47&14.3&0.54&0.37&0.28&0.19&1.88\\
 &1.055&312.9&1.56&15.2 &0.69 &0.45&0.31&0.20&1.79\\
 &1.048&330.8&1.65&16.1&0.77 &0.47&0.33&0.20&1.62\\
 &1.028&353.2&1.75&17.1&0.67 &0.38&0.34&0.19&1.46\\
 \hline
 \multirow{7}*{~2 bar}
 &2.012&257.4&1.29&12.6&0.47 &0.37&0.49&0.38&4.54\\
 &2.037&274.5&1.38&13.4&0.46 &0.34&0.45&0.33&4.18\\
 &2.000&293.2&1.47&14.3&0.40 &0.27&0.37&0.25&3.71\\
 &2.047&312.9&1.56&15.2&0.60 &0.38&0.47&0.30&3.47\\
 &2.050&331.8&1.65&16.1&0.70 &0.42&0.51&0.31&3.15\\
 &2.042&354.8&1.76&17.1&0.64 &0.36&0.40&0.23&2.89\\
 \hline
 \multirow{7}*{~3 bar}
  &3.012&257.1&1.29& 12.6&0.66 &0.51&0.73&0.57&6.80\\
  &2.996&273.2&1.37&13.4&0.65 &0.47&0.63&0.46&6.18\\
  &3.037&295.7&1.48&14.4&0.66 &0.45&0.68&0.46&5.59\\
  &3.050&313.7&1.56&15.3&0.77 &0.49&0.75&0.48&5.15\\
  &3.064&332.4&1.65&16.1&0.98 &0.59&0.89&0.54&4.76\\
  &3.021&354.4&1.76&17.1 &0.75 &0.43&0.61&0.35&4.28\\
 \hline
 \multirow{7}*{~4 bar}
  &4.026&258.1&1.29&12.6&0.98 &0.76&1.07&0.83&9.04\\
  &4.052&274.9&1.34&13.5&0.70 &0.51&0.79&0.57&8.29\\
  &4.048&295.2&1.48&14.4&0.61 &0.41&0.70&0.47&7.46\\
  &4.041&313.1&1.56&15.3&0.73 &0.47&0.78&0.50&6.84\\
  &4.042&332.7&1.65&16.2&0.86 &0.52&0.86&0.52&6.27\\
  &4.000&353.5&1.75&17.1&0.94 &0.54&0.89&0.51&5.68\\
\hline
\hline
\end{tabular}
\end{center}
\end{table*}

\subsection{Measurements: Light scattering in CO$_2$}
\label{measurements}
Measurements of the RB-scattering spectral profile of CO$_2$ gas were
performed for conditions of 0.5 -- 4 bar pressure and temperatures in
the range between 258 and 355~K, as listed in
Table\ \ref{TableCO2GreenConditionsandResults}.
In this table the accurately measured $p$ and $T$
conditions for 29 ($p,T$) measurement combinations are listed as well
as the temperature-dependent transport coefficients: shear viscosity
$\eta_{\rm s}$ and thermal conductivity $\lambda_{\rm th}$. For all
measurements a value for the internal molecular heat capacity of
$C_{\rm int} = 2/2\:R$ is adopted.

\begin{figure*}[htp]
\centering
\includegraphics[scale=0.51]{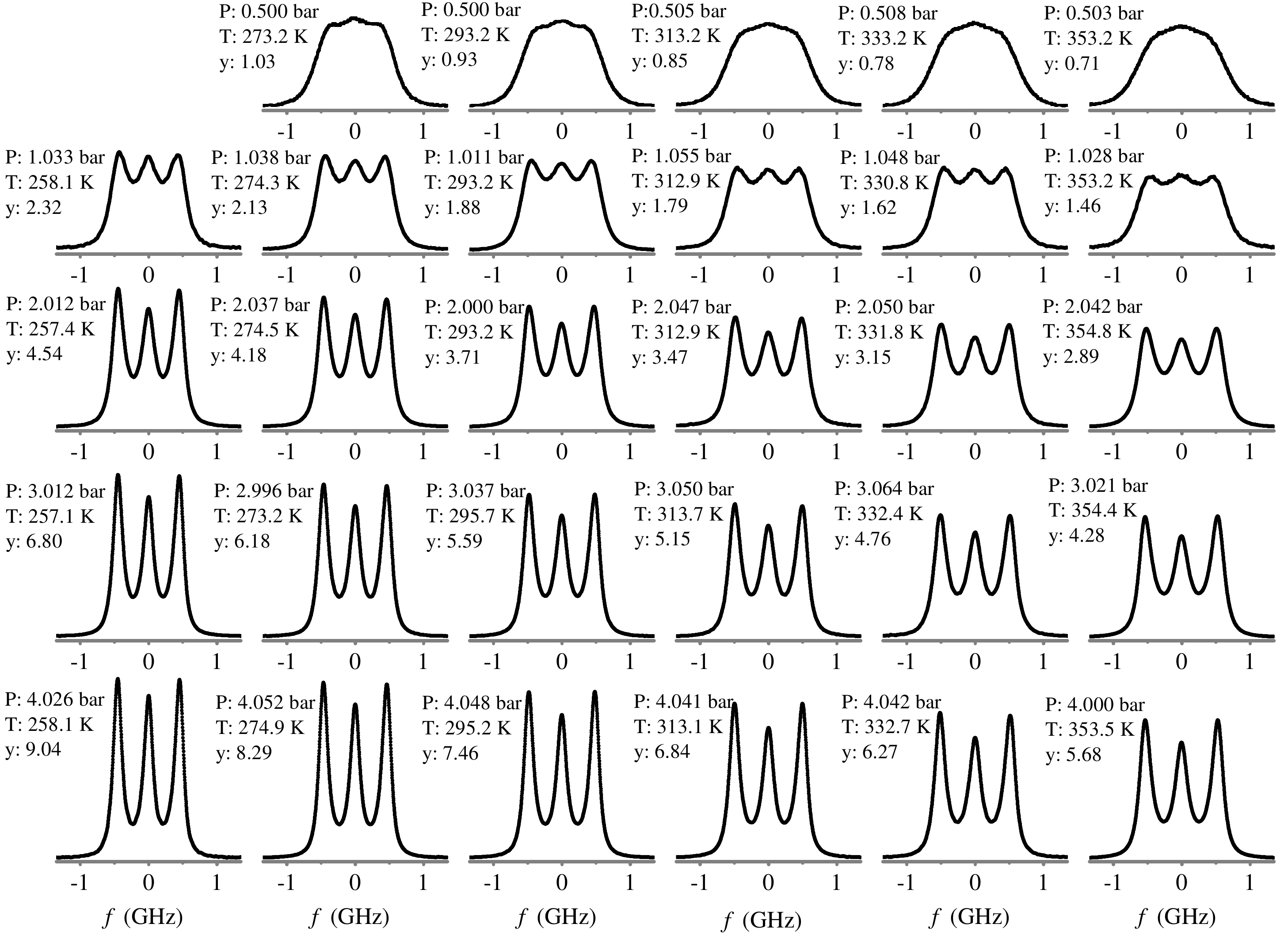}
\caption{Experimental data for RB--scattering in CO$_2$ as measured for the
  various pressure and temperature conditions as indicated. The data is on a scale of normalized integrated intensity over one FSR. The 29
  spectra pertain to the entries in
  Table\ \ref{TableCO2GreenConditionsandResults}.
}
\label{Fig:CO2GreenExpData}
\end{figure*}

In Fig.\ \ref{Fig:CO2GreenExpData} the RB light spectra for the 29
different ($p$,$T$) combinations are graphically displayed. A
qualitative inspection shows, when comparing profiles from the top-
row down, the pressure $p$ is increased and therewith the
$y$-parameter is increased as well, and hence the spectra show more
pronounced Brillouin-side peaks. Indeed at higher uniformity
parameters $y$ the hydrodynamic regime is approached resulting in
well-isolated acoustic side modes. Similarly, while going from left
to right along the columns, the temperature $T$ is increased,
associated with a lowering of the $y$-parameter, and hence the
Brillouin-side peaks become less pronounced. In the following the
experimental profiles will be compared to the Tenti-S6 model and
the Hammond--Wiggins hydrodynamic model.

\subsection{Comparison with the two models}
\label{TentiandHammond--Wiggins}
For a quantitative analysis of the data a comparison is made with the
Tenti-S6 model, that was developed into a code~\cite{Pan2003,Pan2004}
that was included in fitting routines for analyzing both spontaneous
and coherent RB-scattering~\cite{Meijer2010,Vieitez2010}.
In comparing model and experiments, the bulk viscosity $\eta_{\rm b}$
was determined in a least-squares procedure, minimizing the mean squared
deviation
\benn
\label{CO2GreenChisqrue}
  \chi^2 = \frac{1}{N}\sum_{i=1}^{N}
  \frac{[I_e(f_i)-I_m(f_i)]^2}{\delta^2(f_i)}
\eenn
where $I_e(f_i)$ and $I_m(f_i)$ are the experimental and modeled
amplitude of the spectrum at (discrete) frequency $f_i$, and $N$ is
total number of the experimental data.  The error $\delta(f_i)$ of
$I_e(f_i)$ is estimated as the square root of the number of collected
photons.
%

%
First an analysis is made of the spectrum presented in
Fig.\ \ref{Fig:CO2GreenCompareDifferentThermal}. Least-squares fits
are performed for the value of the bulk viscosity $\eta_{\rm b}$
invoking two values for the thermal conductivity, first $\lambda_{\rm
th} = 14.3$ mW/m$\cdot$K as resulting from the modified Eucken
approach and, second the value obtained from a direct measurement at
acoustic frequencies $\lambda_{\rm th}=16.2$ mW/m$\cdot$K
\cite{Bousheri1987}.  The residuals plotted in
Fig.\ \ref{Fig:CO2GreenCompareDifferentThermal} show that in the first
approach a peak residual of 3\% is found, where in the latter
approach the peak deviation amounts to 5\%. This supports validity of
the treatment of thermal conductivity following the modified Eucken
relation in this study conducted at hypersound frequencies. In such
an approach focusing on hypersound the internal heat capacity is set
at $C_{\rm int} = 2/2 \: R$, signifying that two rotational degrees of
freedom are involved and vibrational relaxation is 'frozen'.

\begin{figure*}[t]
\centering
\includegraphics[scale=0.53]{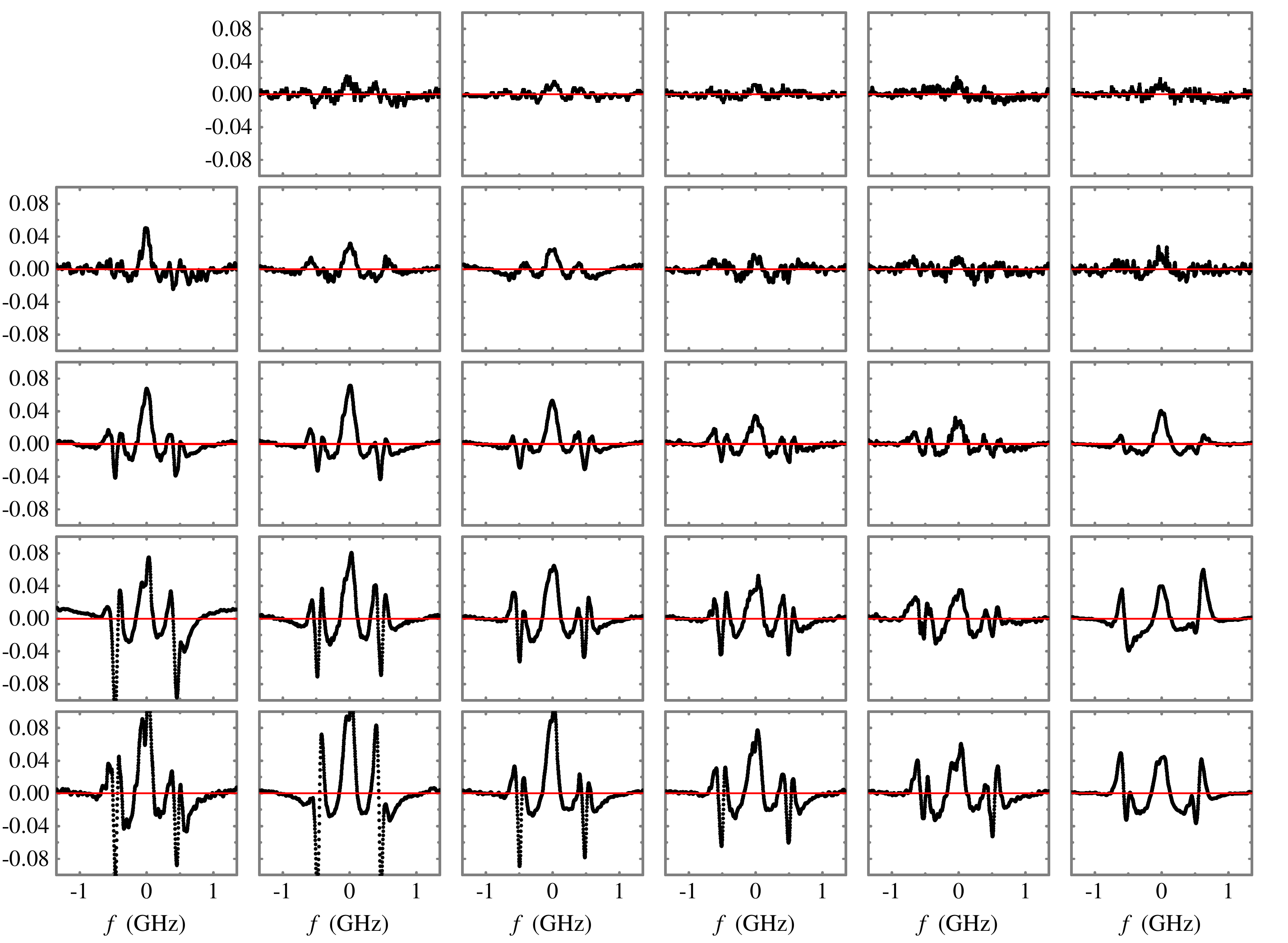}
\caption{Plot of calculated residuals between experimental data and spectral
  profiles obtained from a fit to the Tenti-S6 model with transport
  coefficients as listed in
  Table~\ref{TableCO2GreenConditionsandResults} and optimized values
  for $\eta_{\rm b}$ as derived in the fit. Note the one-to-one
  correspondence with the 29 graphs of spectra in
  Fig.\ \ref{Fig:CO2GreenExpData}.
  }
\label{Fig:CO2GreenTenti-S6Res}
\end{figure*}

Least-squares fitting procedures based on the Tenti-S6 model were
applied to the large body of 29 data sets on CO$_2$ for ($p,T$)
values as displayed in Fig.\ \ref{Fig:CO2GreenExpData}.  With
inclusion of values for the transport coefficients as listed in
Table\ \ref{TableCO2GreenConditionsandResults} optimized values for
$\eta^T_{\rm b}$ were derived. Resulting values are listed in
Table\ \ref{TableCO2GreenConditionsandResults}. Based on these fits
and optimized $\eta^T_{\rm b}$ values, residuals between experimental data
and the Tenti-S6 model description are calculated and displayed in
Fig.\ \ref{Fig:CO2GreenTenti-S6Res}. These residuals provide insight
in the quality of the fit, its accuracy and the applicability of the
model.

\begin{figure*}[t]
\centering
\includegraphics[scale=0.56]{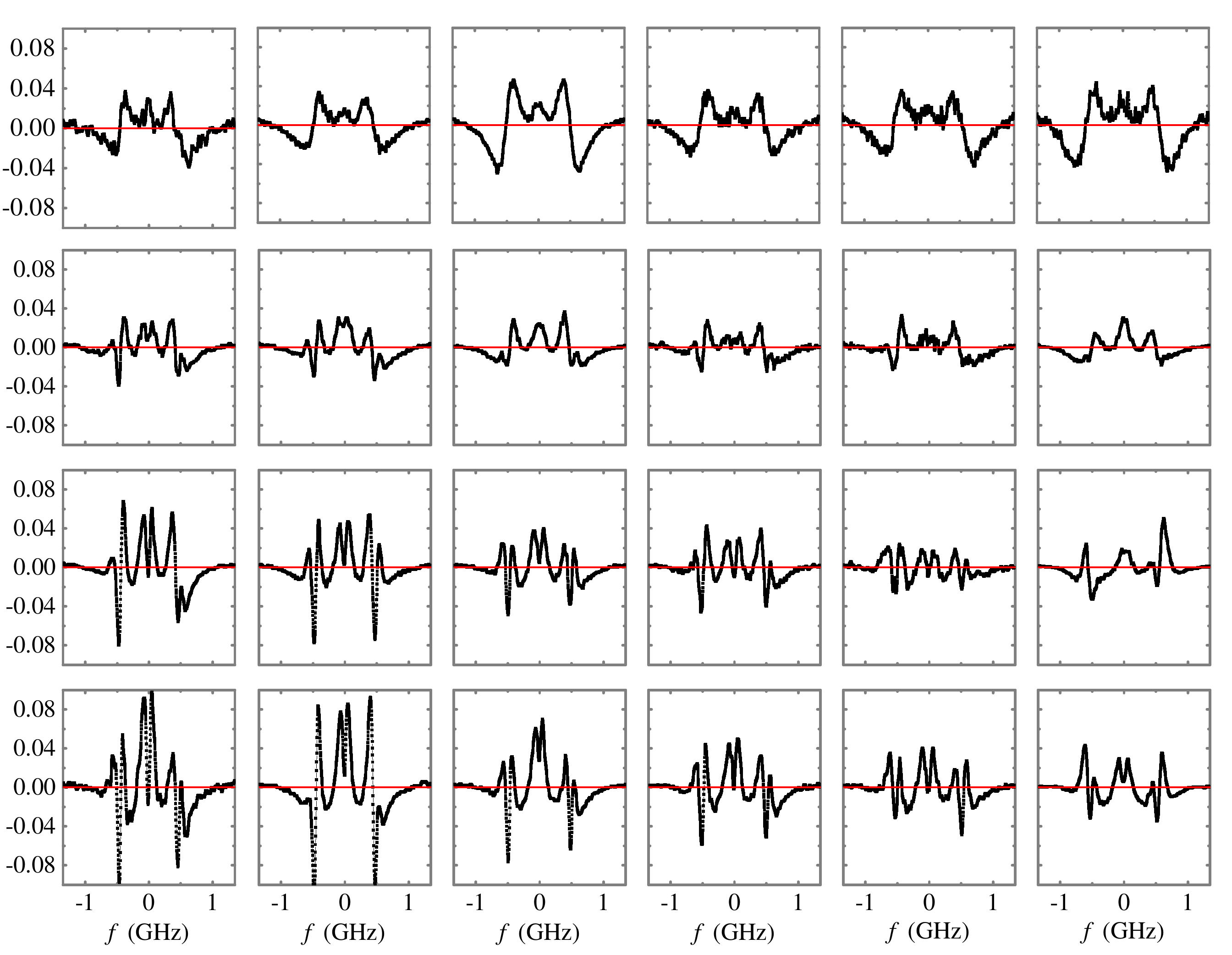}
\caption{Plot of calculated residuals between experimental data and spectral
  profiles obtained from a fit to the Hammond-Wiggins model with
  transport coefficients as listed in
  Table\ \ref{TableCO2GreenConditionsandResults} from 1bar to 4 bar.
  The optimized values for $\tau_{\rm rot}$ as derived in the fit and the
  corresponding $\eta^{H}_{\rm b}$ as derived using Eq.\ (\ref{Eq:ModifiedEuckenSimpleRelax})
  when setting the vibrational relaxational time as 0. Note the
  one-to-one correspondence with the 24 graphs of spectra in
  Fig.\ \ref{Fig:CO2GreenExpData}.
}
\label{Fig:CO2GreenHMRes}
\end{figure*}

Similarly, the hydrodynamic model is used to compare the experimental
and model spectra for the data of pressures above 1 bar. Here, the
rotational relational time $\tau_{\rm rot}$ was adopted as a free
parameter, from which the bulk viscosity was calculated using Eq.\
(\ref{Eq:ModifiedEuckenSimpleRelax}).  Fig.\ \ref{Fig:CO2GreenHMRes} displays the residual
between the experimental data and this model.

\subsection{Bulk viscosities}
\label{sec.bv}

\begin{figure*}[ht]
\centering
\includegraphics[scale=0.48]{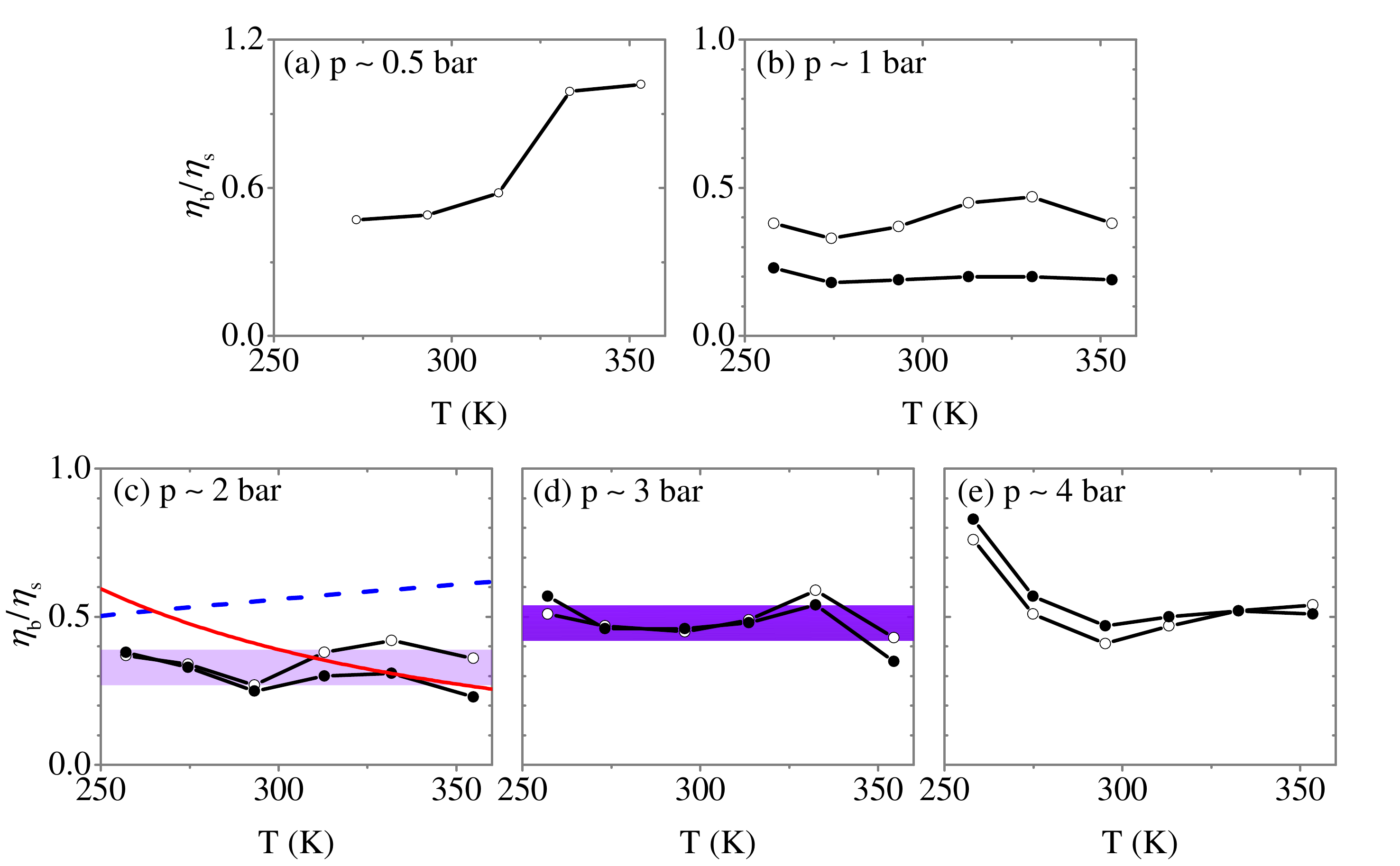}
\caption{
  Summary of results: the ratio of bulk to shear viscosity $\eta_{\rm
  b} / \eta_{\rm s}$ of CO$_2$ as a function of temperature $T$. Open
  circles: as estimated from the experimental data using the
  Tenti-S6 model, closed dots: using the Hammond--Wiggings
  hydrodynamic model.  Red line in (c): Landau--Teller theory {\em
  scaled down by a factor $10^4$}. Blue dashed line in (c):
  prediction of Parker model\cite{Parker1959} for rotational relaxation. Violet bars in
  (c), (d): estimate of mean $\eta_{\rm b} / \eta_{\rm s}$.
}
\label{Fig:CO2GreenRatioofBulktoShear}

\end{figure*}

Our main result, the ratio $\etabs$ as a function of temperature and
pressure is summarized in Fig.\ \ref{Fig:CO2GreenRatioofBulktoShear}. At $p = 0.5 \: {\rm
bar}$ the spectral lineshape is not very sensitive to variation of
$\eta_{\rm b}$, at the highest pressure the models deviate significantly
from the experiment.  At $p = 1\: {\rm bar}$ (corresponding to $y =
[1.4 - 2.3]$), the hydrodynamic model does not yet apply, and
the values of $\etabs$ of the two models differ significantly.

Fig.\ \ref{Fig:CO2GreenRatioofBulktoShear} also shows the prediction of the Landau--Teller
scaling, which captures low-frequency experimental data
\cite{Cramer2012}. However, with a crucial proviso: {\em it is scaled
down by a factor $10^4$}.  This vividly illustrates the dramatic
effect of high frequencies on the ratio $\etabs$.
At these high frequencies, only rotational relaxation remains.  Based on
the analysis of classical trajectories, \citet{Parker1959} derived a
scaling expression for the ratio $\eta_{\rm b}/\eta_{\rm s}$,
\benn
\etabs \propto \left[ 1 + \frac{\pi^{3/2}}{2} \left(
\frac{T^*}{T}\right)^{1/2} +\left(\frac{\pi^2}{4} + \pi \right)
\frac{T^*}{T} \right]^{-1}
\eenn
 with $T^* = 82.6 \: {\rm K}$ the temperature associated with the well depth of the O-O interaction
potential \cite{zhang2005optimized}. This prediction, scaled on $\etabs = 0.5$ at $T = 250\: {\rm K}$, is also shown in Fig.\ \ref{Fig:CO2GreenRatioofBulktoShear}.

We find no significant dependency on pressure or temperature.  At
pressures where both kinetic and hydrodynamic models apply, we find
an average $\etabs = 0.33 \pm 0.06$ at $p = 2 \: {\rm bar}$ and
$\etabs = 0.48 \pm 0.06$ at $p = 3 \: {\rm bar}$.  These averages,
together with their uncertainty, are also indicated in Fig.\
\ref{Fig:CO2GreenRatioofBulktoShear}.
Our present numbers are consistent with the finding obtained from the
light scattering experiments on CO$_2$ in the UV-range ($\lambda =
366.8$ nm) covering the parameter space $y = [0.9 - 3.7]$, and
yielding $\eta_{\rm b} = (5.7 \pm 0.6) \times 10^{-6}$
Pa$\cdot$s~\cite{Gu2014a}, which gives rise to $\etabs = 0.39 \pm 0.04 $.
These values should be compared to $\eta_{\rm b} = 4.6 \times 10^{-6}$
Pa$\cdot$s (for $y = [3.3 - 8.2]$) by \citet{Lao1976a}, corresponding to $\etabs = 0.31 $, and $\eta_{\rm b} =
3.7 \times 10^{-6}$ Pa$\cdot$s for $y = [0.44 - 3.54]$ by
\citet{Pan2005}, for which the ratio $\etabs = 0.25 $.
%
%

\section{Discussion and conclusion}
\label{SecCO2GreenDiscussandConclusion}
In this paper we study Rayleigh-Brillouin scattering over a range of
pressures with the aim of determining the bulk viscosity using two
different types of models for the spectral lineshape.  Where the
range of applicability of these two models overlaps, we find
consistent values of the bulk viscosity.

At low frequencies the bulk viscosity depends strongly on
temperature, which is caused by the temperature dependence of the
vibrational relaxation rate.  We do not find a significant
temperature dependence, not even the one predicted for the increase
of the bulk viscosity with temperature due to the increase of the
{\em rotational} relaxation time.  We find $\etabs = 0.41 \pm 0.10$ at
pressures of 2 and 3 bar.


For N$_2$, a measurement of sound absorption at low pressure yielded
a value of the bulk viscosity, expressed relative to the shear
viscosity as $\etabs = 0.73$ (Ref.\onlinecite{Prangsma1973}). \citet{Cramer2012}
showed that this ratio should increase from 0.4 to 1 as the
temperature changes from 100 K to 420 K. Indeed \citet{Gu2013b}
experimentally determined a ratio of $\etabs = [0.46 - 1.01]$ from
RB-scattering for the temperature range $254.7 - 336.6$~K. Hence, for
measurements at lower sound frequencies ($f_{\rm sl}$) and at hypersound
frequencies ($f_{\rm sh}$), this ratio yields a similar value. The
vibrational relaxation time of N$_2$ is larger than $10^{-4}$ seconds
at room temperature~\cite{Taylor2013}, thus $f_{\rm sl}\tau_{\rm vib} \gg
1$ as well as $f_{\rm sh}\tau_{\rm vib} \gg 1$. In other words, the
vibrational degrees remain frozen under both conditions
\cite{emanuel1990bulk}.

For CO$_2$ a different situation is encountered. At atmospheric
pressure, the vibrational relaxation time is $\tau_{\rm vib} = 6
\times 10^{-6}$ s, while the rotational relaxation time is $\tau_{\rm
rot} = 3.8 \times 10^{-10}$ s (Ref.\ \onlinecite{Lambert1977,Gu2014a}).
Hence, for sound frequency measurements (at the MHz scale), $f_{\rm
sl}\tau_{\rm vib}\thickapprox 1$ and $f_{\rm sl}\tau_{\rm rot}\ll 1$,
which means that both rotation and vibration are excited and take
effect during energy exchange with translation in collisions. The
method of sound absorption delivers an experimental ratio of bulk
viscosity to shear viscosity ($\etabs$) of ${\cal O}(10^{4})$ (Ref.\
\onlinecite{Tisza1942,Prangsma1973,Emmanuel1996}). For hypersound
frequencies (at the GHz scale), $f_{\rm sh} \tau_{\rm vib}\gg 1$,
causing the vibrational modes not to take effect.

In order to describe macroscopic flow phenomena on time scales ranging from microseconds to nanoseconds, the used value of the bulk viscosity will range over four orders of magnitude. An example of such a flow phenomenon is a shock in a high Mach number flow. The comparison of scattered light spectra to kinetic and hydrodynamic models in this paper shows that this dramatic frequency dependence of the bulk viscosity is due to the to the (gradual) cessation of
vibrational relaxation.

\section*{Acknowledgements}

This research was supported by the China Exchange Program jointly run
by the Netherlands Royal Academy of Sciences (KNAW) and the Chinese
Ministry of Education. YW acknowledges support from the Chinese
Scholarship Council (CSC) for his stay at VU Amsterdam. WU
acknowledges the European Research Council for an ERC-Advanced grant
under the European Union's Horizon 2020 research and innovation
programme (grant agreement No 670168).
The core part of the code that computes the Tenti models has been
kindly provided to us by Xingguo Pan.

\newpage

\section*{References}


\end{document}